\documentclass[notitlepage]{article}

\usepackage{authblk}
\usepackage{graphicx}
\usepackage{epstopdf}
\usepackage{subcaption}
\usepackage{amsmath}
\usepackage{multirow}
\usepackage{booktabs}
\usepackage{topcapt}
\usepackage{xcolor}
\usepackage{hyperref}
\usepackage{amssymb}
\usepackage{times}
\usepackage{comment} 

\usepackage{titling}

\newcommand\eg{\emph{e.g.}}
\newcommand\ie{\emph{i.e.}}
\newcommand\etal{\emph{et al.}}

\newcommand{\nop}[1]{}

\usepackage{color}

\begin{document}
\title{Striations in PageRank-Ordered Matrices}
\author{Corey Pennycuff}
\author{Tim Weninger}
\affil{Department of Computer Science and Engineering\\University of Notre Dame}
\date{}                     
\setcounter{Maxaffil}{0}
\renewcommand\Affilfont{\itshape\small}
    \maketitle
    \begin{abstract}
Patterns often appear in a variety of large, real-world networks, and interesting physical phenomena are often explained by network topology as in the case of the bow-tie structure of the World Wide Web, or the small world phenomenon in social networks. The discovery and modelling of such regular patterns has a wide application from disease propagation to financial markets. In this work we describe a newly discovered regularly occurring striation pattern found in the PageRank ordering of adjacency matrices that encode real-world networks. We demonstrate that these striations are the result of well-known graph generation processes resulting in regularities that are manifest in the typical neighborhood distribution. The spectral view explored in this paper encodes a tremendous amount about the explicit and implicit topology of a given network, so we also discuss the interesting network properties, outliers and anomalies that a viewer can determine from a brief look at the re-ordered matrix.
\end{abstract}

\section{Introduction}

Patterns and regularities found in nature frequently inform scientific theories with numerous applications. Regularities within networks give way to the analysis of a rich set of physical phenomenon including viral propagation patterns, social behavior, gene and protein interaction, transportation flow, and so on. These networks are typically understood by computing certain high-level characteristics like the degree and eigenvalue distribution. Deeper analysis of some networks may reveal the presence of community patterns or regularities in the average path length, \ie, small world networks. In these and other cases it is the presence or absence of well-defined patterns that contribute to broadly applicable theories about the nature of the phenomenon present in the network~\cite{Ugander2013}. In many cases, the presence of regular patterns makes the discovery of suspicious or anomalous actors easier leading to improvements in anomaly detection like spam filtering, intrusion and fraud detection, health monitoring, and ecosystem disturbances~\cite{Chakrabarti2004b}.

In this paper we describe the appearance of a newly discovered striation pattern that is present in the PageRank re-ordering of the adjacency matrices of real world networks.

Networks can be easily represented as an \textit{adjacency matrix} $A$, where a cell $A_{ij}$ is set to 1 if nodes $i$ and $j$ are connected by an unweighted edge, or some weight $w$ in the weighted case, and 0 otherwise. If the graph is undirected then the matrix will be symmetric. The number of entries in a row in the adjacency matrix determines the {\em outdegree} of the corresponding node, and the number of entries in the column in the adjacency matrix determines the {\em indegree} of the corresponding node. In an undirected graph their is no concept of in- versus out-degree, therefore only the total {\em degree} of a node is relevant in a symmetric adjacency matrix. Edges define the topology of the network, therefore a visual rendering of the adjacency matrix often leads to meaningful analysis of the network topology~\cite{bertin1973semiologie,kang2011spectral,mueller2007interpreting}. The network topology also informs a wide variety of graph processes that can be used to help rank and order the vertices in the network. One such graph process is random walk, which is a stationary process on undirected graphs, but may not be stationary on directed graphs because of the possible presence of nodes with 0 outdegree, \ie, dangling nodes. In order to make a random walk stationary on a directed graph the random walker must have an opportunity to leave from a dangling end~\cite{Perra2008}.

The PageRank algorithm provides an elegant solution to this problem. In addition to the diffusion process modelled by randomly walking from node to node, PageRank adds a teleportation probability that stochastically jumps the walker to a random node. This function can be described as:

\begin{equation}
\label{eq:pr}
p(i) = \frac{1-d}{n} + d \sum_{j\in A_i}{\frac{p(j)}{k_{\textrm{out}(j)} }},
\end{equation}

where $p(i)$ is the PageRank value of node $i$, $n$ is the number of nodes in the graph, $k_{\textrm{out}(j)}$ is the outdegree of node $j$, and the summation runs over all of the nodes with edges pointing towards $i$. The damping factor $d$ is the teleportation probability that weights the probability of walking an outgoing edge from $i$ or jumping to some other node. Teleportation effectively solves the dangling node problem by moving the walker to a new location when stuck at a dangling node~\cite{page1999pagerank}.

\begin{figure}[t]
    \centering
    \begin{minipage}[t]{4.5in}
        \includegraphics[width=\textwidth]{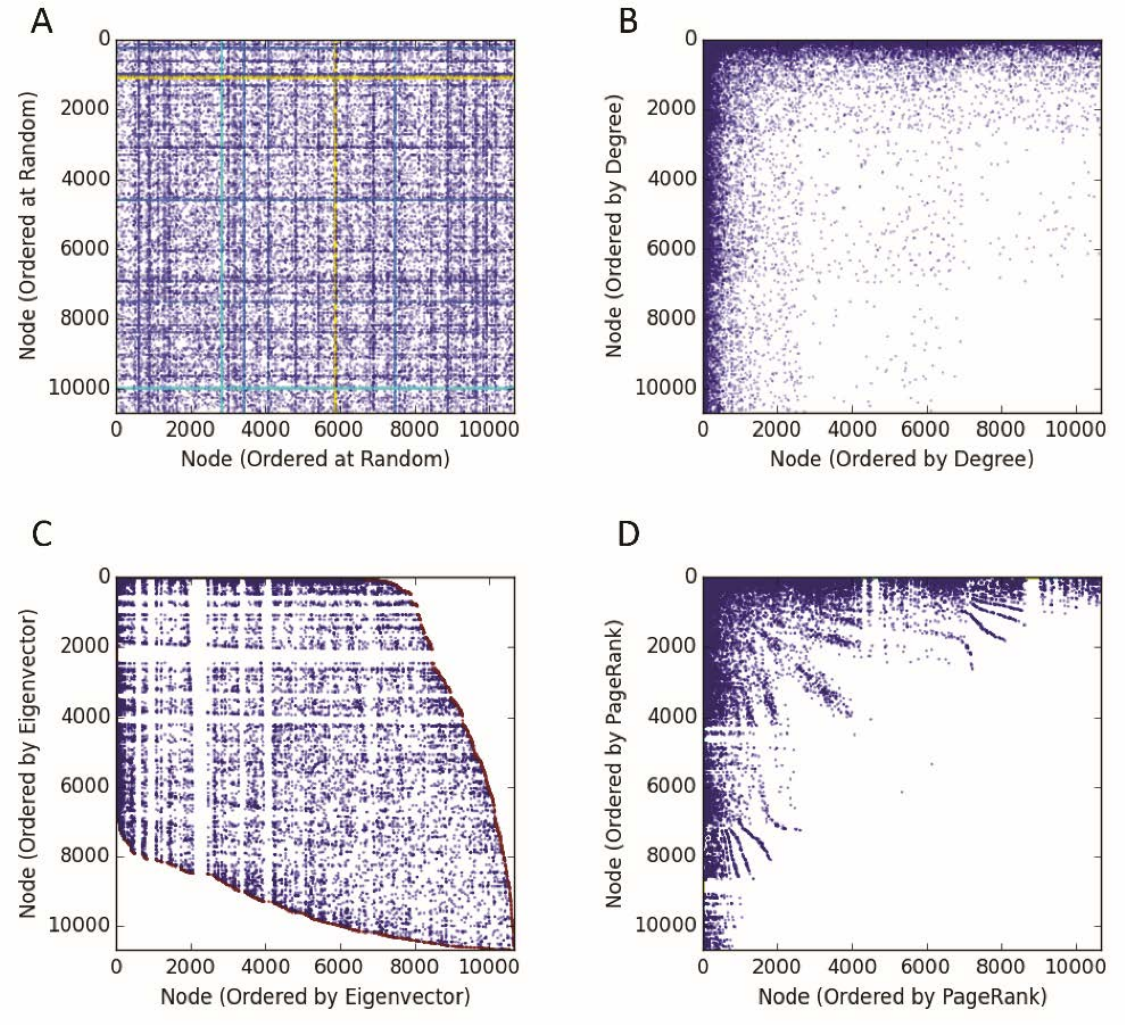}
    \end{minipage}    
    \caption{Adjacency matrices of University of Oregon Internet routing network. Adjacency matrices are ordered by various importance measures: (A) Random, (B) Degree, (C) Eigenvector, (D) PageRank. In (B) the red cells that comprise the outer envelope of the shape denote nodes of degree=1}
    \label{fig:ordering}
\end{figure}

PageRank is related to the principal eigenvector of the adjacency matrix. The eigenvector similarly ranks the importance of a node based on the importance of its incoming neighbors:

\begin{equation}
\label{eq:ev}
\lambda x_i = \sum_{j\in A_i}{x_j} = \sum_{j}{A_{ij} x_j} = ({\textbf{A}^T\textbf{x}})_i, 
\end{equation}

meaning that $x_i$ is the $i^{\textrm{th}}$ component of the eigenvector of the transpose of the adjacency matrix with eigenvalue $\lambda$.

Solving Eq.~\ref{eq:pr} is said to be equivalent to solving Eq.~\ref{eq:ev} on a transition matrix $M$ defined as:

\begin{equation}
\label{eq:gm}
M_{ij} = \frac{1-d}{n} + d \frac{A_{ji}}{k_{\textrm{out}(j)}},
\end{equation}

where $M$ is oftentimes called the {\em Google Matrix}. Thus, the PageRank score $p(i)$ from Eq.~\ref{eq:pr} is often said to be the principal eigenvector of $M$. We typically compute the  eigenvector using power iteration, in which the Google matrix $M$ is repeatedly multiplied by some vector ($p(i)$ in the case of Eq.~\ref{eq:pr} or $x_i$ in the case of Eq.~\ref{eq:ev}) until convergence~\cite{Langville2005}.

Plotting and comparing graphs via PageRank or other importance measures, \eg, degree, closeness, betweenness, has been a foundational representation of networks. Figure~\ref{fig:ordering} shows the adjacency matrix of Internet routing paths~\cite{Leskovec2005} ordered randomly (A), by node-degree (B), by the principal eigenvector score (C) and by PageRank (D) score for each node. 

Random ordering illustrates very little in terms of structural regularity, although the reader may be able to glean the general connectivity (\ie, sparseness/denseness) from the proportion of whitespace. The degree-ordered matrix means that the node with the highest degree is listed at the top, and the edges that connect to other nodes of high degree are denoted with edges appearing towards the left-hand side of the matrix; ties occur frequently in degree-orderings and are broken arbitrarily. From this matrix we can begin to infer that the general degree distribution of the network.

Previously, Chakrabarti et al. discussed the ``water droplet'' pattern formed from adjacency matrices ordered by the principal eigenvector~\cite{Chakrabarti2004b,Chakrabarti2007} (illustrated in Fig.~\ref{fig:ordering}C). They initially found that the eigenvector-ordered adjacency matrix, called the A-plot, ``has a clean and smooth oval-shaped boundary''~\cite{Chakrabarti2004b}. They also found that the boundary corresponds to the 1-degree nodes in the graph; The shape and cleanness of this boundary is explained as follows: if $I_i$ denotes the network value of node $i$ and node $i$ connects only to node $j$, then the properties of spectral decomposition imply

\begin{equation}
\label{eq:wd}
I_i = 1/\lambda_1 \times I_j,
\end{equation}

where $\lambda_1$ is the largest eigenvalue of the adjacency matrix of the graph. Therefore the boundary of edges in the A-plot can be calculated from the first eigenvalue and eigenvector~\cite{Zhan2003}. Eq.~\ref{eq:wd} also shows that there cannot be edges at all outside the boundary (See appendix C of~\cite{Zhan2003} for proof). 

Although not illustrated in Fig.~\ref{fig:ordering}C, there exists a clean and smooth internal boundary that corresponds to 2-degree nodes. Specifically, if node $i$ is connected to two nodes of similar importance values $I_j$, then their edges will be plotted in a fashion similar to the 1-degree nodes:

\begin{equation}
\label{eq:wd2}
I_i = 2/\lambda_1 \times I_j,
\end{equation}

In this same fashion there exist a boundary of 3-degree nodes, etc., but these boundaries are sparse because it becomes vanishingly difficult to find a 3, 4-degree nodes connected to 3 or 4 nodes, respectively, of similar eigenvector centrality.

Adjacency matrix orderings have been used previously to describe the topology of a network. Perhaps the most widely known matrix ordering technique is encompassed by Blockmodelling~\cite{doreian2005generalized} in which a network's adjacency matrix is organized such that edges are grouped into a set of `blocks' that represent coherent clusters, typically situated along the diagonal, from which further downstream tasks and decisions can be improved. Aside from blockmodelling, there are several ways to determine an ordering of a matrix's rows and columns in order to gather some insight into the network structure; a more thorough review of the related literature can be found in Section 2. 

In this paper, we describe newly discovered {\em Striation} patterns that arise from PageRank orderings of the adjacency matrix of unweighted and undirected real world networks. Figure~\ref{fig:ordering}D illustrates one example of these striation patterns on the same real-world Internet routing network that is shown in Figs.~\ref{fig:ordering}A--\ref{fig:ordering}C, the only difference being the ordering of the matrix. Throughout this work we use a wide variety of synthetic and real-world networks to describe the physical process underlying the striking visual patterns that are observed. Ultimately, we find that the shape of the patterns tell an interesting story about the hidden structures (both complex and predictable) that are present in everyday networks.

\section{Related Work}
\label{supp:related}
The origin of PageRank was rooted in the intent to rank web pages based on their link topology~\cite{page1999pagerank}.  Although there are alternative link topology based algorithms such as HITS~\cite{kleinberg1999authoritative} and the SALSA~\cite{lempel2000stochastic} (which combines PageRank and HITS), PageRank enjoys a brand recognition due to its early integration into and association with the Google search engine~\cite{franceschet2011pagerank}. 

Many studies have examined the type and quality of output of the PageRank algorithm.  Page \etal{} advised that $\alpha = .85$ based on empirical evidence~\cite{page1999pagerank}, and Bechetti and Castillo further showed that PageRank does not fit a power-law distribution for extreme $\alpha$ values~\cite{Becchetti2006}.  Pandurangan \etal showed that the PageRank values of the web follow a power law~\cite{pandurangan2002using}, and Volkovich \etal{} showed the correlation between various parameters of the network (in-degree, out-degree, and percentage of dangling nodes) and the overall log-log shape of the PageRank plot~\cite{volkovich2007determining}. 

The adjacency matrix is a common tool for visualizing graph structure, and has been shown to perform well in providing an intuitive understanding of dense or large graphs, except in the task of path finding~\cite{ghoniem2005readability}.  The drawback of matrices, however, is that nodes can appear in any arbitrary order.  Structures can often be revealed by re-ordering the nodes of the matrix to reveal clusters of related nodes.  A study by Mueller \etal{} provided a comparison of different methods of ordering the matrices, using Random, BFS, DFS, King's algorithm, Reverse Cuthill McKee, Degree, Spectral, Separator tree partitioning algorithm, and Sloan ordering~\cite{mueller2007comparison}.  They evaluated the different orderings in terms of their consistency and ability to reveal structure on graphs generated by different algorithms.  They did not, however, investigate PageRank or any of the other ranking-models discussed in the present work. 

The idea of finding order in visual representations of matrices, especially adjacency matrices of graphs, has a long history. McCormick \etal{} introduced the Bond Energy Algorithm~\cite{mccormick1972problem}, which provided a method of reordering the columns and rows of matrices so that larger values were grouped together.  This method used a nearest neighbor heuristic to overcome the $N!$ possible permutations, and is among the earliest algorithms that may be used as a graph clustering algorithm.

One of the simplest graph-centric ordering algorithms is degree ordering, in which the \textit{degree}, \ie, the number of edges connected to a node, is used as a sorting metric.  This is also the most na\"{\i}ve, in that there are usually many ties, in which case the ordering is undefined.

DFS and BFS are both standard and foundational graph theory algorithms.  They both return orderings dependent on the starting node, as well as the arbitrary tie-breaking when choosing an edge to explore or in adding neighboring nodes. Reverse Cuthill McKee (RCM)~\cite{george1971computer,george1981computer} and King's algorithm~\cite{king1970automatic} provide a variation of BFS in which a priority queue chooses which node to visit next.  The RCM priority queue is based on the node's degree, while the King's algorithm priority queue orders the nodes based on how many of the nodes connected to it by its outgoing edges have already been visited.  As in the other methods, ties may occur within the priority queue, and choice of starting node affects the final order.

Sloan's algorithm~\cite{sloan1986algorithm} orders the nodes by defining a start and end node, then prioritizes the remaining nodes by their degree and distance from an end node.  Nodes farthest from the end node and with a low degree have highest priority; as the highest priority nodes are processed, the priority of that node's neighbors is increased, and the cycle continues until all nodes have been processed.  Sloan's algorithm is also dependent on the chosen start and end nodes, and like other variants does not have a global order.

The ordering of the Separator tree is based on an algorithm given by Blandford \etal{}~\cite{blandford2003compact}.  It removes vertices (\textit{separators}) from the graph in order to partition the structure into subgraphs of a fixed size.  Separator trees were a result of trying to reduce the size of storage required for large graphs by isolating and thereby removing the need to store empty parts of the adjacency matrix. 

Mueller \etal, identified common shapes that emerge within ordered adjacency matrices and give explanations of the underlying structure which produces that shape~\cite{mueller2007interpreting}.  They identify diagonal lines, wedges, and blocks.  Diagonal lines do not indicate connected components when the diagonal is only one cell wide, but wider lines identify chains of similarly connected components.  Wedges and blocks indicate fully-connected cliques.  They also identified three common \textit{footprints}, or visual features, that were observed within the various ordered matrices, including \textit{envelopes}, \textit{horizons}, and \textit{galaxies}.  \textit{Envelopes} are a border-type visual pattern that makes the adjacency matrix resemble a leaf; the edge, or envelope, gives the appearance of a solid border around the leaf shape.  A \textit{horizon} is an area along the diagonal of the adjacency matrix that is more dense than the surrounding area.  A dense \textit{horizon} may indicate chains of interconnected nodes.  A \textit{galaxy} is a sparse distribution of edges within a graph with no discernible structure.  It is commonly observed in a randomly or poorly ordered adjacency matrix.  A \textit{galaxy} does not imply that there is no structure, but rather that the current ordering does not reveal any structure. The primary thrust of their work was to show that the idea of adjacency matrices could be extended to Visual Similarity Matrices, with which one could quickly observe the similarity in structure of various graphs. Mueller did not, however, identify the striation patterns of PageRank orderings demonstrated in the present work.  

The traditional adjacency matrix is not the only method available to tease patterns from graph topology.  Prakash \etal{}~\cite{prakash2010eigenspokes} found a pattern called EigenSpokes, which is observed by creating a scatter plot of the singular vectors (an \textit{EE-plot}) of the nodes against each other upon the presence of an edge.  This plotting method often forms lines that align along specific vectors (\textit{spokes}), which is then used in community detection.  Kang \etal{}~\cite{kang2011spectral} demonstrated that nodes which have high scores on the EE-plot will often form near-cliques or bipartite cores.

\begin{figure}[t!]
    \centering
    \begin{minipage}[t]{.99\textwidth}
        \includegraphics[width=\textwidth]{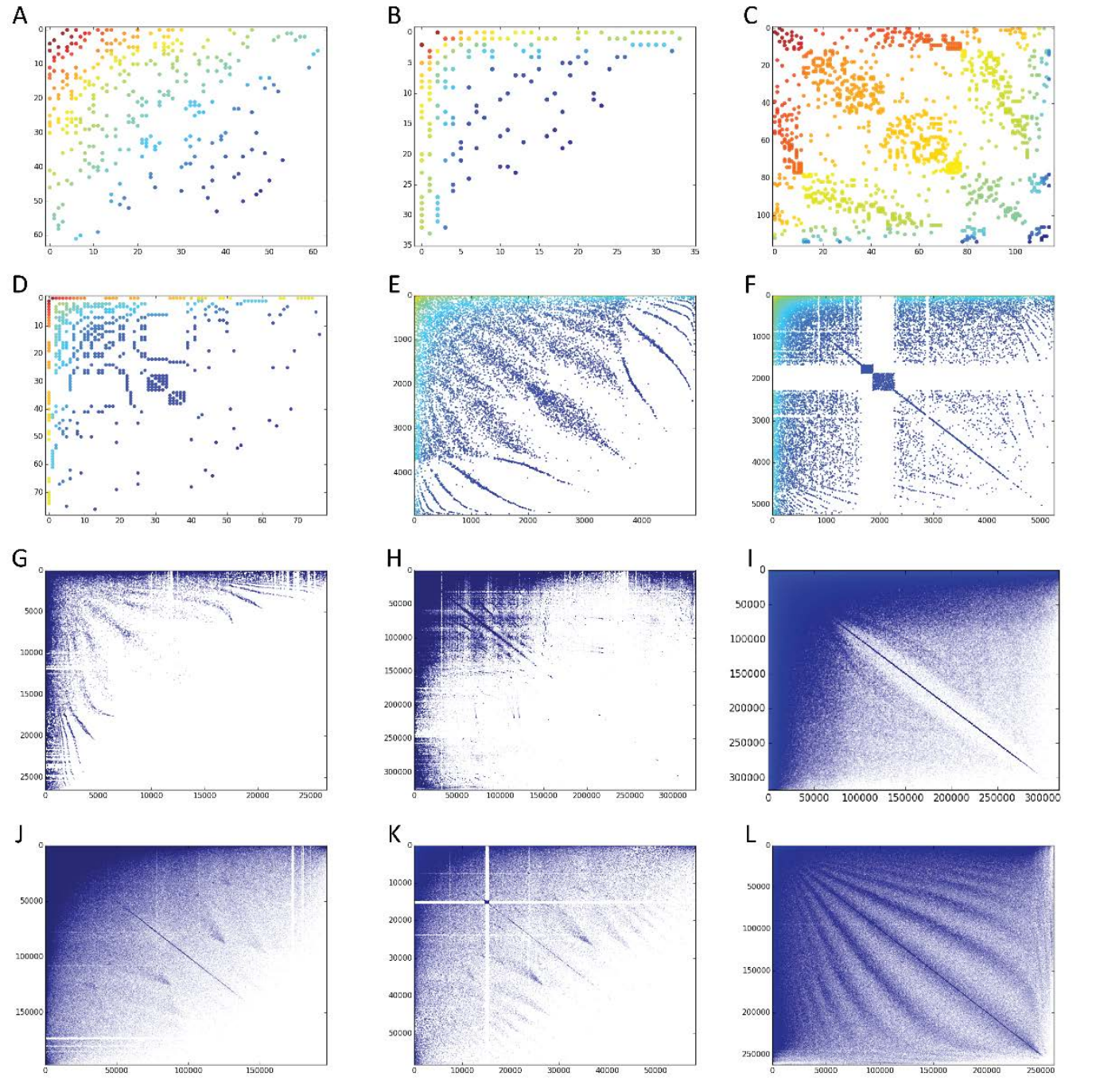}
    \end{minipage}%
    \caption{PageRank ordered adjacency matrices of real world networks: (A) Dolphins, (B) Karate, (C) Football, (D) Les Miserables, (E) Power Grid, (F) ArXiv Relativity and Cosmology Collaborations, (G) CAIDA, (H) Notre Dame Web Site, (I) DBLP Collaborations, (J) Gowalla Social Network, (K) Brightkite Social Network, (L) Amazon Co-Purchases. Each dot/cell represents and edge connecting two nodes. Cell-color denotes the sum of connected nodes' PageRank score.}
    \label{fig:realworld}
\end{figure}

The present work is the first to explore the PageRank-ordering of adjacency matrices of large networks. The next section presents the findings and explains the striation patterns as an inherent result of two fundamental properties of real-world networks.

\section{Results}

We begin by illustrating the PageRank orderings of 12 real-world networks of various sizes and types. Figure~\ref{fig:realworld} shows that the striation patterns tend to originate from the high-PageRank vertices located in the top-right of each matrix downwards toward the lower ranked vertices at the bottom and towards the right. Whitespace in the adjacency matrix, denoting the absence of edges, is clearly correlated with lower-ranked vertices -- an unsurprising finding because vertices with high PageRank also tend to have a high degree.

The adjacency matrices illustrated in Fig.~\ref{fig:realworld} are listed from smallest to largest, but we notice other differences in the size and shape of the various matrix patterns. For example, the football (\ref{fig:realworld}C) and Amazon (\ref{fig:realworld}L) networks have relatively few broad striations that are maintained throughout the entire matrix. On the other hand, the CAIDA (\ref{fig:realworld}G), Power Grid (\ref{fig:realworld}E) and, to a lesser extend, the Notre Dame Web Site (\ref{fig:realworld}H), the Arxiv collaboration network (\ref{fig:realworld}F) and the Les Miserables co-scene network (\ref{fig:realworld}D) have thin striations that do not emanate throughout the entire matrix; these same matrices are also noticeably sparser in their low-ranking region than the former networks.

\begin{figure}[t!]
        \includegraphics[width=.99\textwidth]{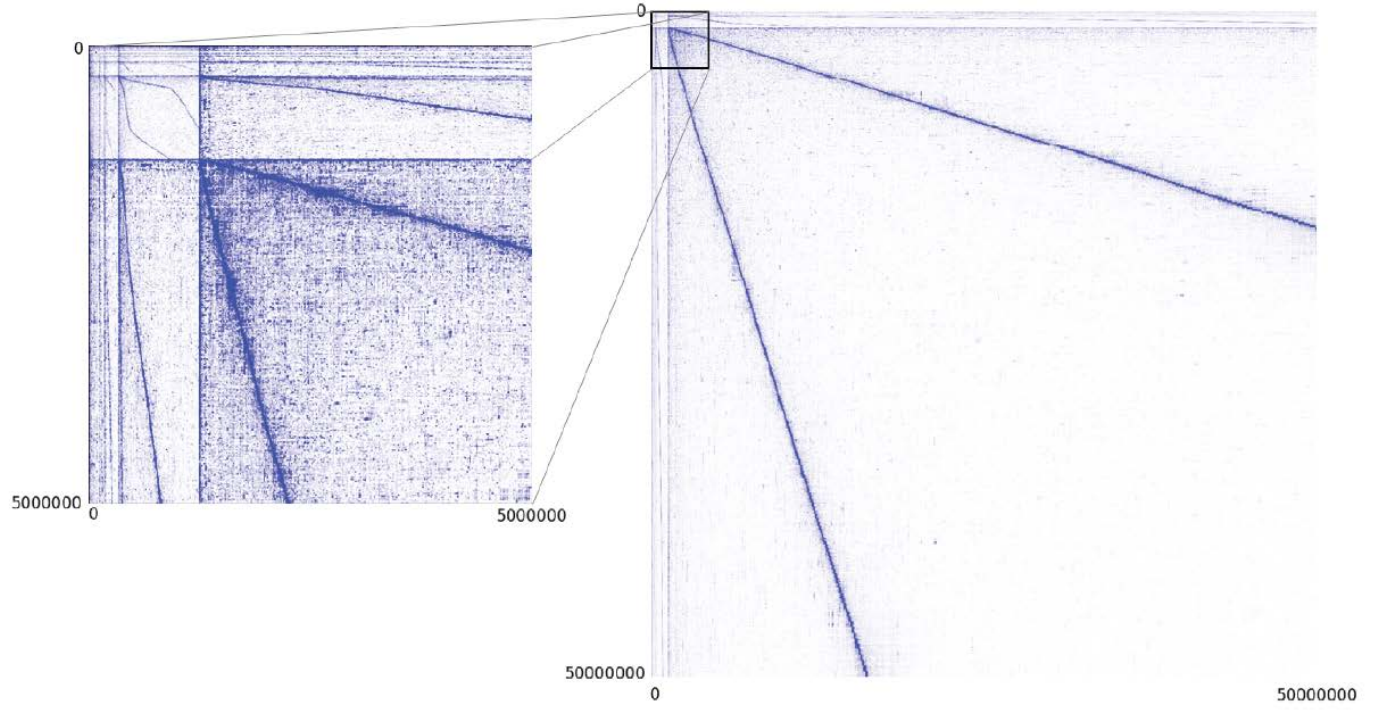}
        \caption{PageRank ordered adjacency matrices of the undirected ClueWeb B data set if 50 million nodes. Each dot/cell represents and edge connecting two nodes. }
        \label{fig:500K}
\end{figure}

Our largest network, the Clueweb crawl of English Web pages in 2009, contains full information of about 50 million nodes. Due to the tremendous size of the resulting adjacency matrix, we are limited in our ability to render the entire matrix. We were, however, able to partially render 1\%-sized subsets of the matrix and stitch the 10,000 sub-matrices together to resolve the full adjacency matrix. It is difficult to view certain detail in the overall matrix, so Fig.~\ref{fig:500K} also shows the first 10\% sub-matrix (left) in addition to the full matrix (right); we find the same striation pattern here as we see in the other adjacency matrices. 

Throughout the remainder of this section we explore these and other differences using synthetic network simulations and point to known qualities of the real-world graphs in order to infer the cause of the various striation types.

\subsection{Generating Scale-Free Striations}

\begin{figure}[t]
    \centering
    \begin{minipage}[t]{4.5in}
        \includegraphics[width=\textwidth]{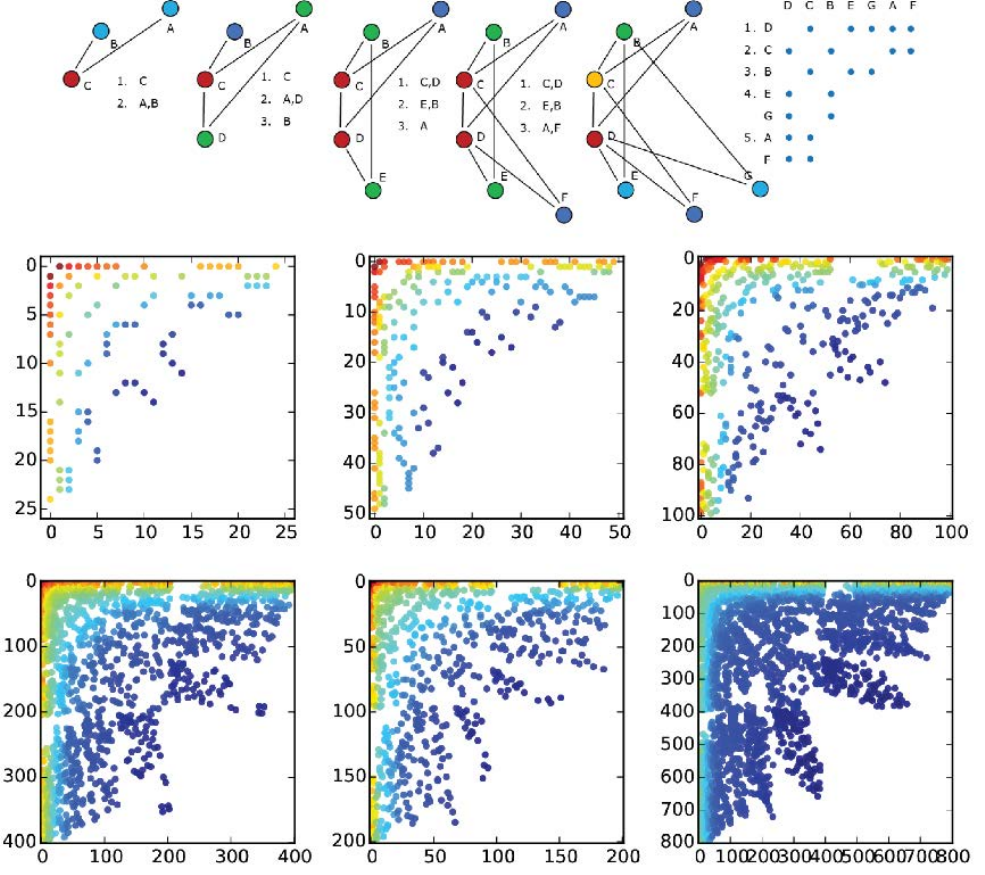}
    \end{minipage}%
    \caption{Scale Free network generation by preferential attachment $\gamma=3$. Top row illustrates the growth model where and the initial adjacency matrix of the right-most network containing 7 vertices. Each dot/cell represents and edge connecting two nodes. Cell-color denotes the sum of connected nodes' PageRank score. The bottom row shows the continued growth of the network via preferential attachment 25, 50, 100, 200, 400 and 800 nodes. This generative model clearly grows networks with a certain striation pattern. Although preferential attachment is a stochastic model, repeated runs generate very similar PageRank-ordered adjacency matrices.}
    \label{fig:growing_ba}
\end{figure}

To investigate this phenomenon further we grow undirected graphs using preferential attachment~\cite{Liben2007} as illustrated in Fig.~\ref{fig:growing_ba} with $\gamma=3$. Each graph iteration is ordered and colored by their PageRank scores. The top row of this figure shows how the preferential attachment process grows the graph from steps 4 through 7 corresponding to graphs with sizes of $n=4$ through $n=7$. The PageRank rankings for the nodes in each graph demonstrate how symmetry arises from the graph generation model. The explicit PageRank ordered matrix in the bottom row of Fig.~\ref{fig:growing_ba} shows the early stages of what grow to be striation patterns. Further growth is illustrated in the six graph snapshots of size $n=25$, $50$, $100$, $200$, $400$, and $800$ respectively. We also generated networks up to $n=$1,000,000 and found that the striation patterns remain consistent within a given $\gamma$ parameter.

\begin{figure}[t]

 \centering
    \begin{minipage}[t]{4.5in}
        \includegraphics[width=\textwidth]{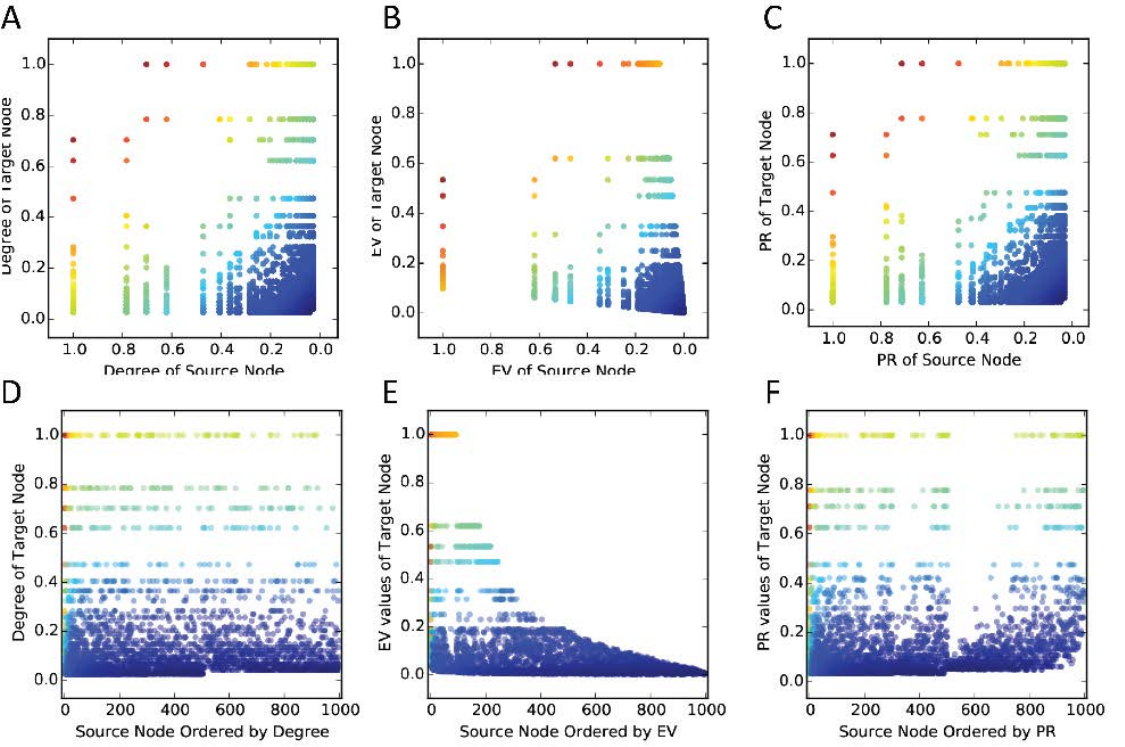}
    \end{minipage}%
  \caption{Degree (A,D), eigenvector centrality (B,E) and PageRank (C,F) values by edge attachments. Top row plots value association; bottom row plots value-to-order association. $Y$-axis indicate values, normalized between 0 and 1, and are consistent for each row. All figures plot exact same network generated by preferential attachment ($n=1000$, $\gamma=3$).}
  \label{fig:value_ordering} 
\end{figure}

We are further interested in the explicit causes of the striation patterns that are apparently caused by regularities emanating from the generative process. As demonstrated in Figure~\ref{fig:value_ordering}C and in related literature, we frequently find that PageRank~\cite{Becchetti2006} and eigenvector centrality~\cite{Chakrabarti2004} of individual nodes follow a power law distribution. These findings reflect the power laws found in degree distributions of real world graphs~\cite{Barabasi1999} due to the inherent relationship between PageRank, eigenvector centrality and degree.



This investigation led us to plot the PageRank-values and orderings of the edges in the network. Figure~\ref{fig:value_ordering} plots the degree (\ref{fig:value_ordering}A and~\ref{fig:value_ordering}D), eigenvector centrality (\ref{fig:value_ordering}B and \ref{fig:value_ordering}E) and PageRank values (\ref{fig:value_ordering}C and \ref{fig:value_ordering}F) of edges in the top row, and the value-to-ordering association in the bottom row. The top row illustrates how the various importance values of the source nodes are associated with the values of the nodes it connects to. Figure \ref{fig:value_ordering}A indicates that nodes of high degree (normalized from 0 to 1) are attached to other nodes of various degree; the same can be said for the other value-to-value plots in Fig.~\ref{fig:value_ordering}B and \ref{fig:value_ordering}C. The top row equates to a two-dimensional plot expressing two facets of graphs generated by preferential attachment: 

\begin{enumerate}
\item PageRank scores follow a power law distribution~\cite{Becchetti2006}, and 
\item PageRank values of a typical node's neighborhood also follow a power law distribution.
\end{enumerate}

These two facets result in the striation patterns demonstrated above. To understand why, consider the bottom row of Fig.~\ref{fig:value_ordering} which plots the source nodes by their ranked-order rather than their values, essentially stretching the plots in the top row across the x-axis. Stretching the value-to-value plots on the top row into value-to-ranking plots on the bottom row show a half-way-there illustration of the striations; if we were to stretch the y-axis in the same way as the x-axis the resultant plots would be the order-by-order plots that show the striations in Fig.~\ref{fig:ordering}B--\ref{fig:ordering}D according to their respective orderings.

Recall that the top row of Fig.~\ref{fig:value_ordering} indicates the values of each node and its neighbor. The matrix is symmetric because the graph is undirected, therefore each source-to-target attachment will have a symmetric attachment in the opposite direction. The degree, eigenvector and PageRank values are spaced as expected due to the well known power law distribution of these three importance measures. 

The major difference between Figs.~\ref{fig:value_ordering}E and \ref{fig:value_ordering}F indicate two interesting dynamics in the attachment properties of eigenvector values and PageRank values. First, high values of eigenvector centrality tend to exclusively attach to nodes of a high value (therefore a high ordering), while the values of PageRank attachments are not distributed as cleanly.

We will investigate the neighborhood distributions more explicitly in Section~\ref{sec:nodenbr}, but first we investigate the Watts-Strogatz process~\cite{Watts1998}, \ie, the small world network generator, to demonstrate how differences in network generation result in different striation patterns. Different settings may help explain the underlying generation process of the real world networks.

\subsection{Generating Small World Striations.}

\begin{figure}[t!]
    \centering
    \begin{minipage}[t]{4.5in}
        \includegraphics[width=\textwidth]{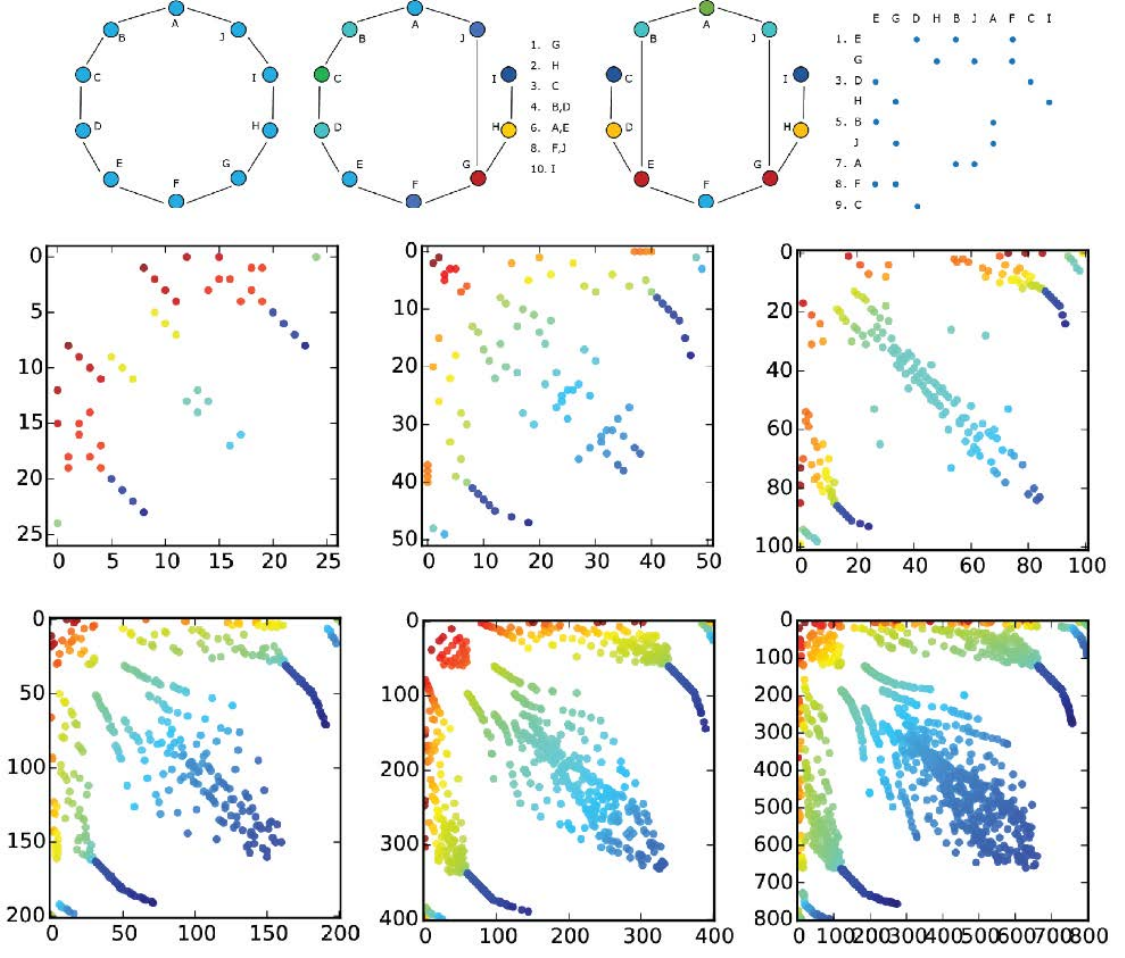}
    \end{minipage}%
    \caption{Small World network generation by Watts-Strogatz process ($n=1000$, $k=12$, and $p=0.20$). Top row illustrates the growth model and the initial PageRank ordered adjacency matrix of the right-most network containing 10 vertices; color represents relative PageRank score. The bottom row shows the continued growth of the network via the Watts-Strogatz process at 25, 50, 100, 200, 400 and 800 nodes. This generative model clearly grows networks with a certain striation pattern. Although Watts-Strogatz is a stochastic model, repeated runs generate very similar PageRank-ordered adjacency matrices.}
    \label{fig:growing_ws}
\end{figure}

Aside from preferential attachment, the small world networks of Watts and Strogatz~\cite{Watts1998} show distinctly different striation patterns caused by their growth model. The Watts-Strogatz process initially creates a ring where each node is attached to $k$ neighbors symmetrically. From this ring an edge $e=[u,v]$ has a probability to reattach to some random node $x$, basically reassigning the end point of some node to $e=[u,x]$. Two iterations of this reassignment process is captured at the top of Fig.~\ref{fig:growing_ws}. This process creates networks with a Poisson degree distribution.  The PageRank ordered matrix of the top-center network is illustrated at the top-right and shows that the structural regularity of the small world network create the beginnings of a striation pattern. The example on the top row is purposefully drawn symmetrically, but the remaining six graph snapshots of sizes $n=25$, $50$, $100$, $200$, $400$, and $800$ (with an initial ring of 2 neighbors each) have a random rewiring probability of 20\% and can rewrite a given edge to any other node. The larger graphs are rather unrealistic because they are created with an initial neighborhood ring of $k=2$, whereas the small world model typically calls for a neighborhood ring of size $k\gg\ln(n)$. More realistic graphs and real world graphs are investigated below. 

During ordering, PageRank ties are broken arbitrarily, but we find that there are rarely ties in the resulting values, therefore the patterns are not a result of tiebreaking. Instead, we posit that the striation patterns are the result of the growth mechanisms on undirected graphs due to the structural regularity that these synthetic growth mechanisms produce.


\begin{figure}[t]
    \centering
    \begin{minipage}[t]{4.5in}
        \includegraphics[width=\textwidth]{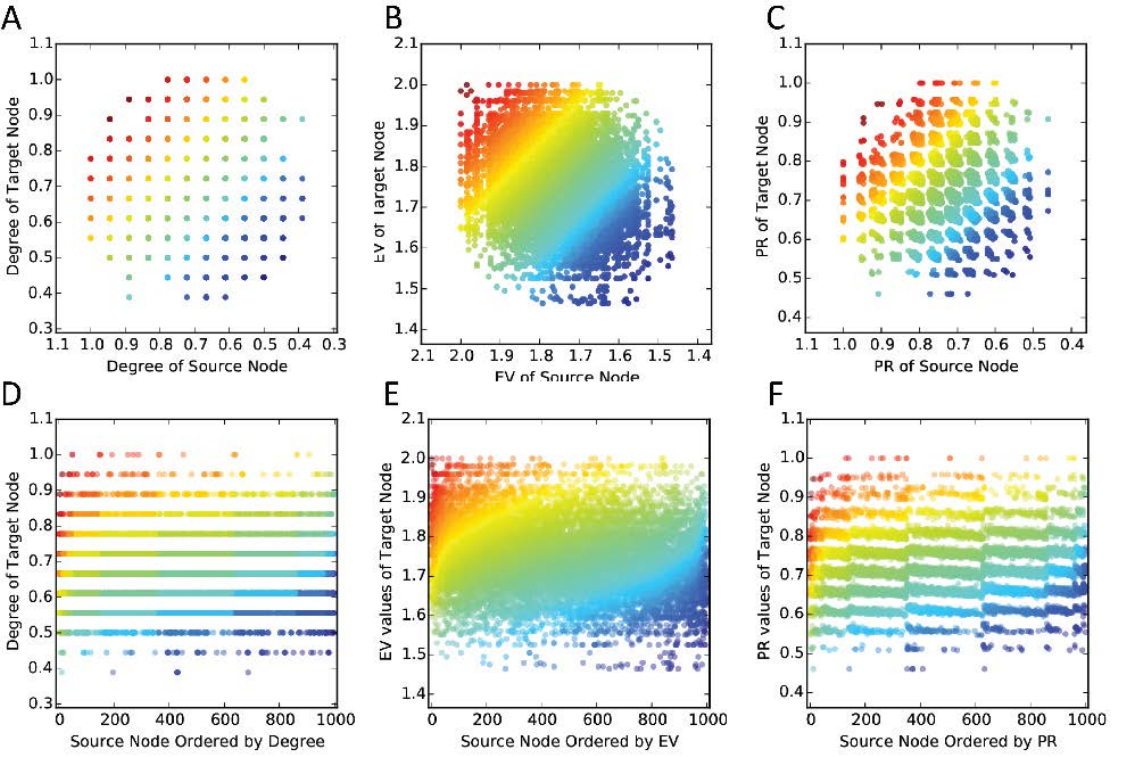}
    \end{minipage}
  \caption{Degree (A,D), eigenvector centrality (B,E) and PageRank (C,F) values by edge attachments. Top row plots value association; bottom row plots value-to-order association. Y-axis indicate values, normalized between 0 and 1, and are consistent for each row. All figures plot exact same network generated by the Watts-Strogatz process ($n=1000$, $k=12$, and $p=0.20$). }\label{fig:sm_value_ordering}
\end{figure}

Figure~\ref{fig:sm_value_ordering} shows the same progression of figures for the Watts-Strogatz process (which results in Poisson distributions for degree, eigenvector and PageRank values) as Fig.~\ref{fig:value_ordering} does for the preferential attachment process (which results in power law distributions for degree, eigenvector and PageRank values). The top row shows value-by-value plots for degree, eigenvector and PageRank values. The bottom row shows corresponding value-by-rank plots where the x-axis is stretched by the ordering. Just as before, if we were to stretch the y-axis of the bottom row, then the plots from Fig.~\ref{fig:growing_ws} would appear.

The degree plots in Fig.~\ref{fig:sm_value_ordering}A~and~\ref{fig:sm_value_ordering}D show the Poisson distribution in discrete steps; this represents the distribution of whole-number degrees normalized between 0 and 1. Therefore each point in Fig.~\ref{fig:sm_value_ordering}A represents several overlapping edges of the same value (with a higher density towards the center), which are fanned out horizontally in \ref{fig:sm_value_ordering}D. The eigenvector plots in Figs.~\ref{fig:sm_value_ordering}B~and~\ref{fig:sm_value_ordering}E also show a Poisson distribution, but with with far fewer ties.

The PageRank results in Figs.~\ref{fig:sm_value_ordering}C~and~\ref{fig:sm_value_ordering}F are particularly interesting. Figure~\ref{fig:sm_value_ordering}C shows that the PageRank distribution is a Poisson. However, we also find that the distribution is not continuous, \ie, there are, in this case, 12 groupings of PageRank-densities across both axes. When stretched by ordering in Fig.~\ref{fig:sm_value_ordering}F we can see that the density of edge-points moves from the top-left to the bottom-right in a slightly diagonal fashion within each of the $12\times 12=144$ mini-sectors. When ordered along the y-axis, these top-left to bottom-right patterns become the striations present in Fig.~\ref{fig:growing_ws} where the density in the value plots are translated into area in the ordered plots.

By varying the small world generation parameters we can see a how the striation patterns change. Changes to $k$ while keeping $p$ invariant generally affects the number of clusters. The number of clusters generally grows with $k$, but is not necessarily equal.

\begin{figure}[t!]
    \centering
    \begin{minipage}[t]{\textwidth}
        \includegraphics[width=\textwidth]{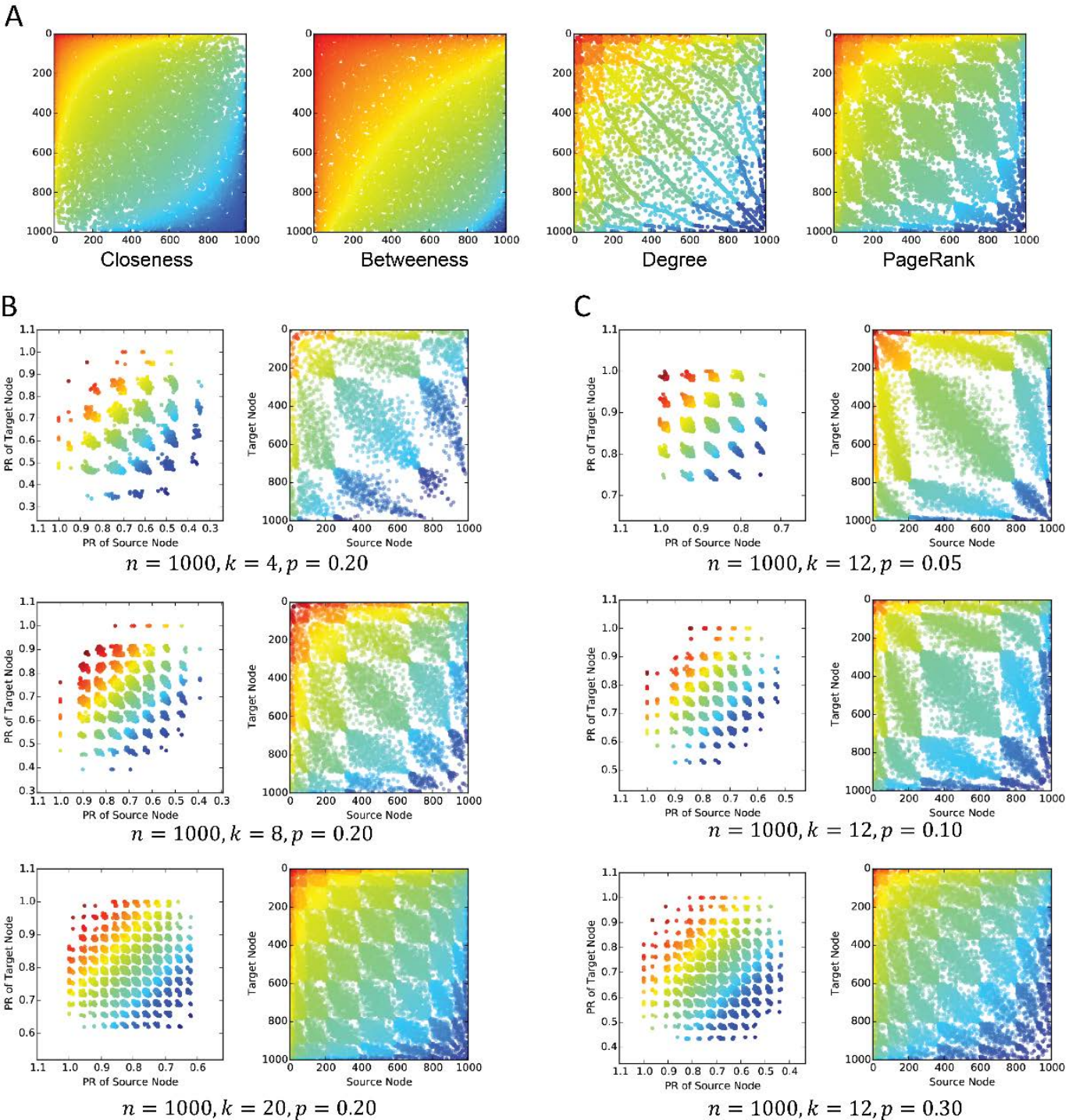}
    \end{minipage}
    \caption{Various Plots of Watts Strogatz Graphs. (A) Adjacency matrix of 1000-node small world graph generated by Watts-Strogatz process. Matrices are ordered by various importance measures. (B) Value-by-value and Order-by-Order PageRank adjacency matrices generated by 1000 node Watts-Strogatz model as p=0.20 and k varies. (C) Value-by-value and Order-by-Order PageRank adjacency matrices generated by 1000 node Watts-Strogatz model as $p$ varies and $k=12$}
    \label{fig:sm_ordering}
\end{figure}

\subsection{Variations on the Watts-Strogatz Process} 
As with preferential attachment, the Watts-Strogatz process also produces a structural regularity within generated graphs. Figure~\ref{fig:sm_ordering}A shows the result of ordering a small world graph with $n=1000$, $k=12$, and $p=0.20$. The closeness and betweenness plots do not exhibit any particular pattern, but rather a simple value gradient. The degree plot in Fig.~\ref{fig:sm_ordering}A draws clear lines representing tie-breaking; and the PageRank ordered matrix shown in Fig.~\ref{fig:sm_ordering}A shows a striation pattern similar to the striations in the preferential attachment graphs, where, instead of a power law distribution, these gaps clearly resemble the Poisson.

By varying the small world generation parameters we can see a how the striation patterns change. Changes to $k$ while keeping $p$ invariant generally affected the number of clusters. Although, $k=12$ typically resulted in 12 clusters this is likely a coincidence; the number of clusters generally grows with $k$, but is not necessarily equal. Figure~\ref{fig:sm_ordering}B shows the plots as $k$ varies from $4$ to $8$ to $20$. In these plots, and in other experiments (not shown) the number of clusters tends to track with $k$, but not exactly. However, the size of the center clusters shrinks consistently as $k$ grows. As $k$ approaches $n$, \eg, $k=n/10$, $k=n/5$, clusters are no longer evident because the graph becomes near completely connected.

Figure~\ref{fig:sm_ordering}C shows the adjacency matrix plots as $k=12$ and the rewiring probability $p$ varies. The number of clusters increases and the size of each cluster decreases as the rewiring probability increases. 

It is difficult to show all potential pairs as $p$ and $k$ vary together. In general, we find that a low $k$ value results in few, large clusters regardless of $p$. Similarly we find that, in general, a low $p$ value also results in few, large clusters.

\subsection{Node Neighborhoods} 
\label{sec:nodenbr}
Here we discuss why these striation-patterns appear in the PageRank ordered matrices. It has been previously observed that PageRank distributions of Web graphs tend to follow a power law distribution~\cite{pandurangan2002using}. We have also observed that, in directed graphs, the in-degree of a node and the its PageRank are highly correlated, meaning that the in-degree distribution of a graph also follows a power law distribution with the same exponent $\alpha$. Generally speaking, this means that the probability that the PageRank and/or in-degree of a node takes a value $x$ is approximately proportional to $x^{-\alpha}$. Aside from Web data, undirected uses of PageRank abound in related work~\cite{abbassi2007recommender,andersen2006local,Perra2008}, where the (undirected) degree and PageRank distributions are correlated in a variety of data sets, not just Web data. For example, if an undirected graph has a power law degree distribution, as in the case of Web data and other scale free networks, then the PageRank is very likely to also have a power law distribution; likewise, a graph with a Poisson degree distribution, as found in Small World graphs~\cite{Barrat2000}, is very likely to have a Poisson PageRank distribution.

\begin{figure}[t!]
    \centering
    \begin{minipage}[t]{4.5in}
        \includegraphics[width=\textwidth]{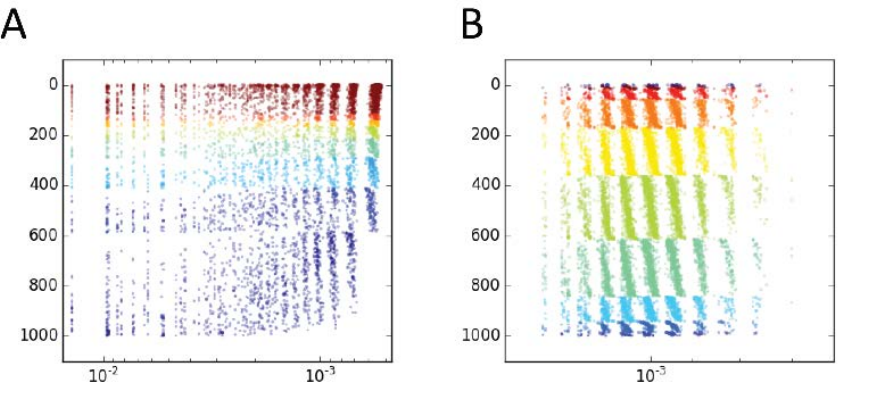}
    \end{minipage}
\caption{PageRank distributions of local neighborhoods in (A) a random network generated by preferential attachment ($n=1000$, $\gamma=3$), and (B) a random network generated by the Watts-Strogatz process ($n=1000$, $k=14$, $p=.20$). Color represents degree of the node, where each node occupies a point on the $y$-axis ordered by PageRank.}
    \label{fig:synth_nbr}
\end{figure}

With degree and PageRank distributions in mind, we ask: what does the degree and PageRank distribution of a single node's neighborhood look like?

Previous work has found that the neighborhoods of individual nodes share similar characteristics as the overall graph. Work in spam detection, for example, finds that if the PageRank distribution of some Web page does not match the overall graph's PageRank distribution, then that Web page is an anomaly and therefore likely to be spam. Furthermore, strong findings on self-similarity and assortativity further indicate that the PageRank in a neighborhood should have the same statistical properties as in the overall graph~\cite{barabasi2000scale,Newman2002}. The similarity between global PageRank and neighborhood PageRank distributions is the key to the presence of striation patterns.

Figure~\ref{fig:synth_nbr} shows the PageRank distribution of each nodes's local neighborhood on the x-axis. Note that the x-axis in the preferential attachment network (\ref{fig:synth_nbr}A) uses a log-scale, which, combined with the edge densities, indicates a power-law distribution in most node-neighborhoods. Similarly, the x-axis in the Watts-Strogatz network (\ref{fig:synth_nbr}B) uses a linear-scale, which, combined with the edge densities, indicates a Poisson distribution present in most node-neighborhoods~\cite{Barrat2000}.

Each row in Fig.~\ref{fig:synth_nbr} shows a nodes' neighborhood PageRank distribution. For example, each source node (\ie, each row in the matrix) has $k$ neighbors drawn from a power law (Fig.~\ref{fig:synth_nbr}A) or Poisson distribution (Fig.~\ref{fig:synth_nbr}B), and these neighbors have PageRank values drawn from a the same distribution.

\begin{figure}[t!]
    \centering
    \begin{minipage}[t]{\textwidth}
        \includegraphics[width=\textwidth]{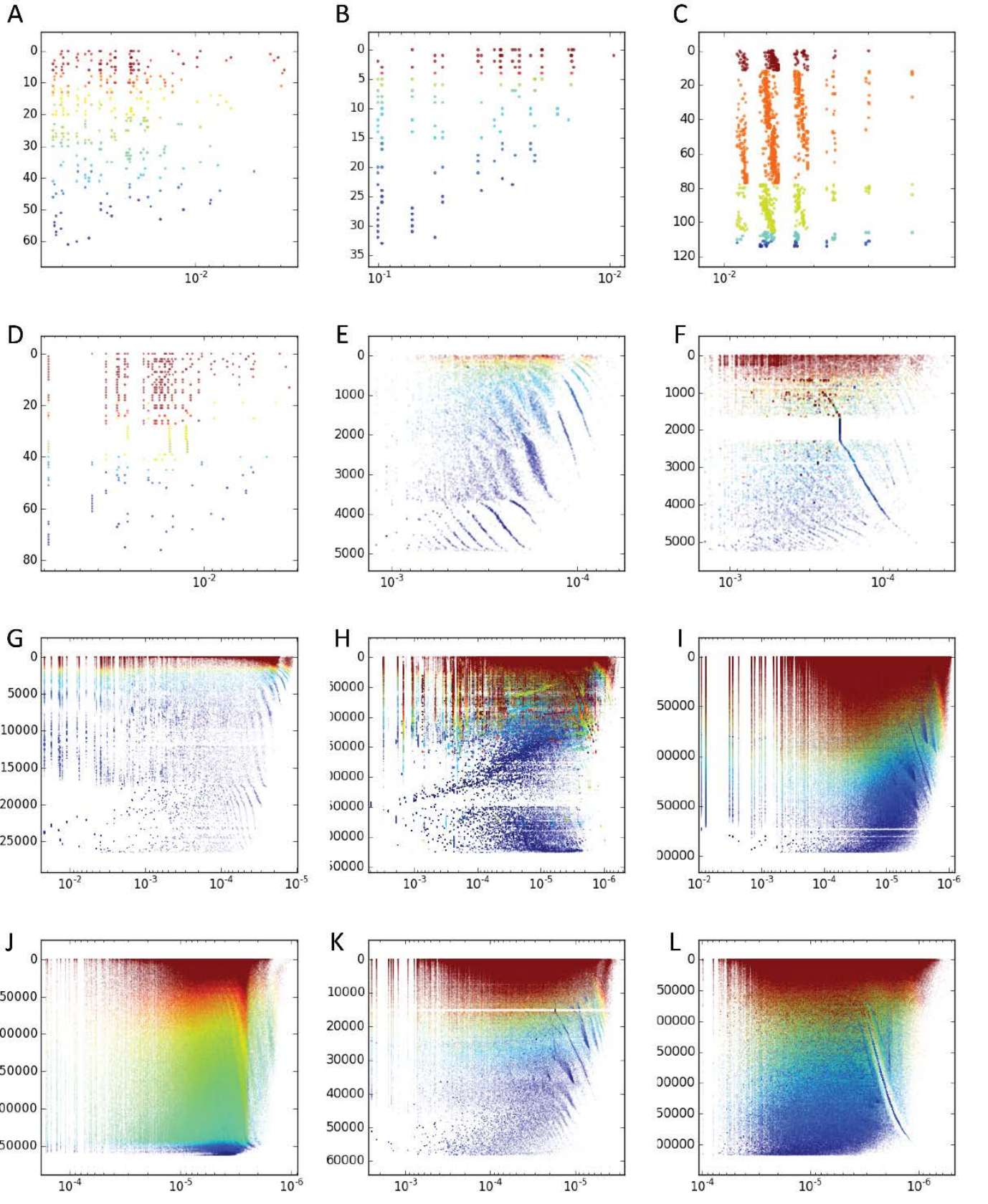}
    \end{minipage}
    \caption{PageRank ordered adjacency matrices of real world networks (A) Dolphins, (B) Karate, (C) Football, (D) Les Miserables, (E) Power Grid, (F) ArXiv Relativity and Cosmology Collaborations, (G) CAIDA, (H) Notre Dame Web Site, (I) DBLP Collaborations, (J) Gowalla Social Network, (K) Brightkite Social Network, (L) Amazon Co-Purchases. X-axis in log-scale. Cell color represents degree of the node (dark-red indicate degree $\ge$ 10), where each node occupies a point on the y-axis ordered by PageRank.}
    \label{fig:realworld_nbr}
\end{figure}

Taken together, we find that striation patterns tend to emerge from similarities in the PageRank-distribution of vertices and the PageRank-distribution of the typical node's neighborhood. Preferential attachment creates networks whose global PageRank-distributions and typical node-neighborhood PageRank distributions follow a power law. Likewise, the Watts-Strogatz process creates networks with nodes and node-neighborhoods whose PageRank distributions follow Poisson distributions. 

To reiterate the key point, striations appear in the PageRank-ordered adjacency matrix when the typical neighborhood's PageRank distribution matches the global PageRank distribution.

The striation patterns are a direct consequence of the orderings on these self-similar distributions. Due to the connectivity mechanics of PageRank (and eigenvector centrality) low-valued nodes are connected to by other low valued nodes. Within the generative process the second lowest-valued node will be be connected to a neighborhood with a slightly higher overall PageRank than the lowest node, and so on. The gradual increase in PageRank on the x-axis, combined with similar PageRank distributions across the y-axis create the striations apparent in these plots.

\begin{figure}[t]
    \centering
    \begin{minipage}[t]{\textwidth}
        \includegraphics[width=\textwidth]{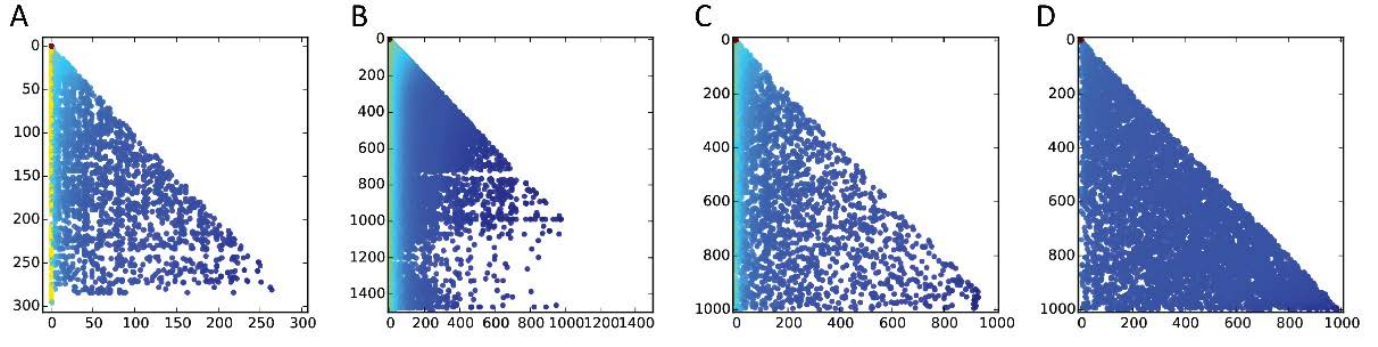}
    \end{minipage}%
    \caption{PageRank orderings of directed networks (A) C. Elegans, (B) Political Blogs, (C) directed preferential attachment network, (D) directed Watts Strogatz network, do not exhibit striation patterns.}
    \label{fig:directed}
\end{figure}

The neighborhoods of the real world networks originally illustrated in Fig.~\ref{fig:realworld} are shown in Fig.~\ref{fig:realworld_nbr}. Here we find that the neighborhood distributions are not as clearly defined as in the synthetic networks of Fig.~\ref{fig:synth_nbr}, nevertheless, certain properties of the synthetic networks can be applied to understand the properties of the real-world networks. For example, the Football network (\ref{fig:realworld_nbr}C, on a log-scaled x-axis) appears to have neighborhoods with a Poisson PageRank distribution resembling the Watts-Strogatz networks from Fig.~\ref{fig:synth_nbr}B. The CAIDA Internet routing network (\ref{fig:realworld_nbr}G) has a power law neighborhood distribution very similar to the preferential attachment network from Fig.~\ref{fig:synth_nbr}A. Other neighborhood plots show similarities with one, both or neither of the synthetic generative processes indicating avenues for further research.

\subsection{Directed Networks} 

Here we look at some instances where striations do not appear. The previous graphs have all been undirected, as evident by their matrix-symmetry. Directed graphs do not exhibit the same striation patterns found in the undirected graphs shown above. 

\begin{figure}[t]
    \centering
    \begin{minipage}[t]{3.5in}
        \includegraphics[width=\textwidth]{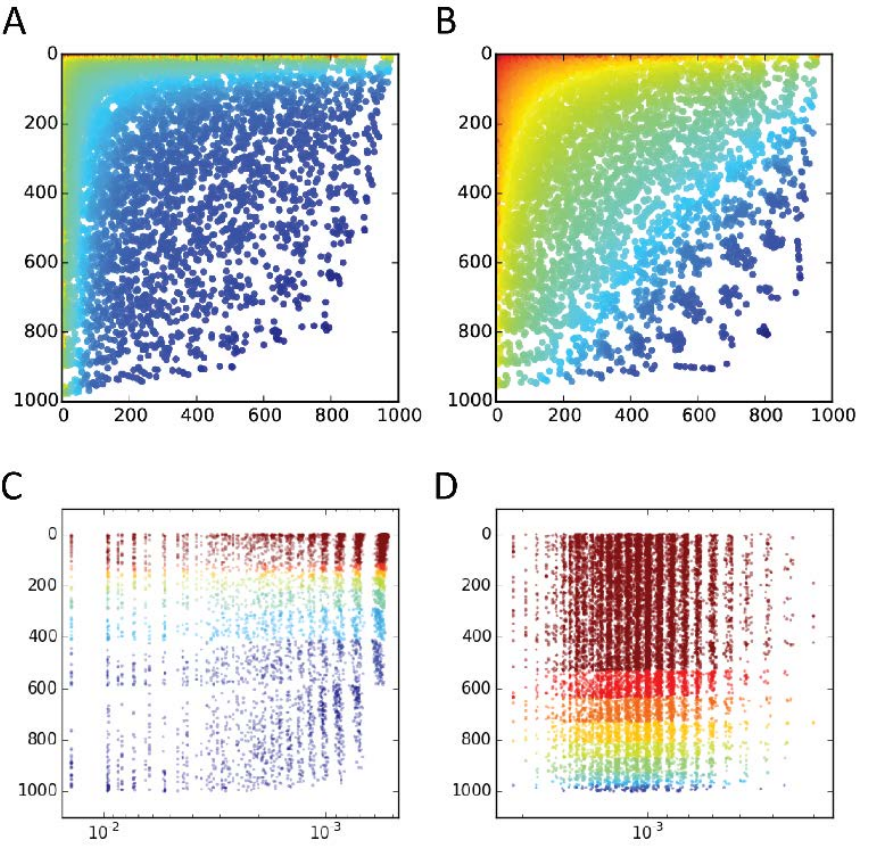}
    \end{minipage}
    \caption{PageRank orderings of undirected networks generated at random to conform to the scale free degree and Poisson distributions of preferential attachment (A) and Watts Strogatz (B) graphs respectively (at top). Cell-color denotes the sum of connected nodes' PageRank score, and each dot/cell represents and edge connecting two nodes. Corresponding neighborhood plots (C and D) are illustrated below; where color corresponds to x-axis node degree.}
    \label{fig:degseq}
\end{figure}

Figure~\ref{fig:directed} illustrates PageRank ordered matrices from two real-world networks (A and B) and two (C and D) generated networks. The preferential attachment graph in Fig.~\ref{fig:directed}C was created with $n=1000$ and $\gamma=3$. The Watts-Strogatz (WS) graph in Fig.~\ref{fig:directed}D was created as an undirected graph with $n=1000$, $k=12$ and $\beta=.20$. After creation the graphs were converted into a directed graph using a fair coin flip to decide the directionality of each edge. Erdos-Reyni random graphs and many others, not shown, similarly do not exhibit striations in both the undirected and directed cases.

During ordering, PageRank ties are broken arbitrarily, but we find that there are rarely ties in the resulting values. Therefore the patterns are not a result of tiebreaking.

Clearly, the striation patterns are the result of the processes resulting in topology regularities that result in PageRank distributions unique in the undirected case only. We are unsure why the striation patterns do not appear in directed networks. It is important to remember that the damping factor in the PageRank algorithm was introduced to deal with the problem of sink-nodes, wherein a non-jumping random walker would get "stuck" in a sink node. Undirected networks, on the other hand, do not have sink-nodes because the random walker can exit any node by the link from which it arrived. Future work might explore this topic further by slowly introducing backward-edges to a directed graph to see if and when striation patterns emerge.

\subsection{Random Networks} 

To explore the causality of the growth mechanism we fed a random graph generator a power-law degree sequence and a Poisson degree sequence representing the results of the preferential attachment and Watts-Strogatz mechanisms respectively. Rather than using these well defined graph processes, we can also generate random graphs where edges are randomly connected with the constraint that the final degree distribution matches some required distribution. Figure~\ref{fig:degseq} shows the PageRank ordered matrix of graphs randomly generated to conform to power law and Poisson degree sequences. We find that striation patterns are largely absent from the randomly generated graphs corresponding to a much smoother, \ie, more random distribution of edges in the neighborhood plot.

\begin{figure}[t!]
    \centering
    \begin{minipage}[t]{\textwidth}
        \includegraphics[width=\textwidth]{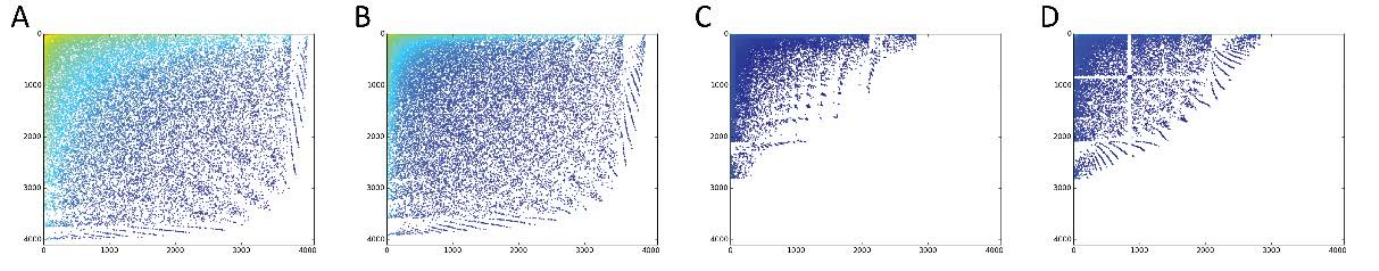}
    \end{minipage}
    \caption{ PageRank ordered adjacency matrices of Kronecker approximations of real world graphs (A) ArXiv Relativity and Cosmology collaborations, (B) DBLP Collaborations, (C) Notre Dame Web Site, (D) CAIDA.}
    \label{fig:kronapprox}
\end{figure}

\subsection{Kronecker Approximations} 
Kronecker graphs are a new approach to modeling real-world networks as $n\times n$ adjacency matrices, where $n$ is relatively small, \ie, typically between 2 and 4. The Kronecker graph model is based on a recursive construction that uses the Kronecker product $\otimes$ to multiply the initial $n\times n$ adjacency matrix by itself recursively~\cite{Weichsel1962}. Leskovec \etal found efficient ways to learn the weights of the initiator matrix in order to recursively generate full-sized graphs that match some real world matrix. For example, the Notre Dame Web site matrix from Fig.~\ref{fig:kronapprox}C is generated by recursively taking the Kronecker product of the initiator matrix $[[0.999,0.414],[0.453,0.229]]$ 12 times resulting in an adjacency matrix with $2^{12}=4096$ rows and columns. The values of the initiator matrix are estimated by fitting an initiator matrix to the real-world network~\cite{Leskovec2010}.

\section{Methods}
\begin{table}[t]
\centering
\begin{tabular}{r|l|l|l|l}
  Network  &  Vertices & Edges &Figure                      &       Source                      \\ \hline
 Dolphins & 62 & 159 & Fig. 2A        &      \cite{Lusseau2003}$^\dag$ \\
 Karate & 34 & 78 & Fig.2B          &       \cite{Zachary1977}$^\dag$ \\
 Football & 115 & 613 & Fig. 2C        &       \cite{Girvan2002}$^\dag$ \\
 Les Miserables & 77 & 254 & Fig. 2D          &       \cite{Knuth1993}$^\dag$ \\
 Power Grid & 4,941 & 6,594 & Fig. 2E           &       \cite{Watts1998}$^\dag$ \\ 
 ArXiv Rel. \& Cos. Collab. & 5,242	& 14,496 & Fig. 2F         & \cite{Leskovec2007}$^\ddag$\\
 CAIDA & 6,474 & 13,233& Fig.  2G    &   \cite{Leskovec2005}$^\ddag$\\
 Notre Dame Web Site & 25,729 &1,497,134 & Fig. 2H       & \cite{Albert1999}$^\ddag$\\
 DBLP Collaborations & 317,080 & 1,049,866 & Fig. 2I    &       \cite{Yang2015}$^\ddag$\\
 Gowalla Social Network & 196,591 & 950,327 & Fig. 2J       &   \cite{Cho2011}$^\ddag$\\
 Brightkite Social Network & 58,228 & 214,078 & Fig. 2K    &   \cite{Cho2011}$^\ddag$\\
 Amazon Co-Purchases & 262,111 & 1,234,877 & Fig. 2L          &   \cite{Leskovec2007b}$^\ddag$\\
 Clueweb2009 B & 428,136,614 & 454,075,638 & Fig. 3        &   $\ast$\\
\end{tabular}
\caption{Sources of real networks from Fig. 2 and 3. Networks denoted with $^\dag$ were downloaded from Mark Newman's network data collection \url{http://www-personal.umich.edu/~mejn/netdata/}. Networks denoted with $^\ddag$ were downloaded from the SNAP data collection \url{http://snap.stanford.edu/data/}. $\ast$ The ClueWeb B graph is available from \url{http://lemurproject.org/clueweb09/}.}
\label{tab:data}
\end{table}

Table~\ref{tab:data} describes the various data sets used in Fig.~\ref{fig:realworld}, Fig.~\ref{fig:realworld_nbr}, and throughout this paper. Directed or multigraph networks were converted to undirected, simple networks if needed. All edge weights, timestamps and other metadata (if any) were ignored. Networks sizes ranged from 34 vertices and 78 edges in the Karate network to 325,729 vertices and 1,497,134 edges in the Notre Dame network. Synthetic networks were created using the NetworkX toolkit~\cite{Hagberg2008}. PageRank, eigenvector centrality and other measures were calculated using NetworkX and confirmed with matlab, numpy and scipy implementations.

\section{Discussion}

We have shown the presence of certain striation patterns found in PageRank ordered adjacency matrices of synthetic and real-world networks. These striations arise in networks that are associated with the well known PageRank distributions combined with similar, regular neighborhood distributions. For example, if a network with a power law PageRank distribution also has vertices with neighborhoods that typically exhibit a power law PageRank distribution, then the resulting PageRank-ordered matrix will exhibit striations. 

The total processing time is bounded by the time it takes to compute PageRank, sort the PageRank result and plot the edges in the graph. To compute the PageRank we use the power iteration method, which has a polynomial time complexity in the size of the matrix. In all cases we limited the number of power iterations to at most 50, however many datasets reach convergence before 50 iterations. Except in the ClueWeb dataset, the largest graphs' PageRank scores could be computed, sorted and plotted in less than 5 minutes on a modern laptop. Once loaded into memory, the 50 million node ClueWeb subset took about an two hours for the submatrices to be computed and plotted and stitched together on a compute machine with 1TB of RAM.

The resulting plots provide a deep look into the topology of a given network, and the observed patterns may be helpful to researchers and practitioners when analysing large and complex networks, especially in the search for anomalous edges, vertices or whole networks. It is our intent to raise more questions that answer, and we solicit the communities help to further understand these patterns. We are working to find a closed form solution describing the striation patterns akin to Eq.~\ref{eq:wd} and Eq.~\ref{eq:wd2}, although the results do not suggest that a strong boundary exists in PageRank-ordered matrices like those present in eigenvector centrality-ordered matrices.


\end{document}